\documentclass[a4paper,11pt]{article}
\pdfoutput=1
\usepackage{jheppub}
\usepackage{amsmath,amssymb,bm,graphicx,bbold,epsf,colordvi}
\usepackage{lipsum}
\usepackage{makecell}
\usepackage{soul}
\usepackage{braket}
\usepackage{bm}
\usepackage{appendix}
\allowdisplaybreaks 
\usepackage{color}
\usepackage[dvipsnames]{xcolor}

\usepackage{mathrsfs}
\usepackage{slashed}
\allowdisplaybreaks
\usepackage[abs]{overpic}
\usepackage[export]{adjustbox}
\usepackage{bm}
\usepackage{scalerel}
\usepackage{accents}

\newcommand{\bef}{\begin{figure}[hbt]\centering}
\newcommand{\eef}{\end{figure}}
\usepackage{mathtools}
\usepackage{subfigure}
\usepackage{booktabs}
\usepackage{graphicx}
\graphicspath{ {./figure/} }
\usepackage{comment}
\usepackage{lipsum}

\newcommand{\beq}{\begin{equation}}
\newcommand{\eeq}{\end{equation}}
\def\bea#1\eea{\begin{align}#1\end{align}}

\def \be  {\begin{equation}}
\def \ee  {\end{equation}}
\def \ba  {\begin{eqnarray}}
\def \ea  {\end{eqnarray}}

\allowdisplaybreaks

\makeatletter
\def\@fpheader{~}
\makeatother

\usepackage[Q=yes,pverb-linebreak=no]{examplep}

\title{Azimuthal decorrelation for photon induced dijet production in ultra-peripheral collisions of heavy ions}

\author[a]{Cheng Zhang}
\author[a]{, Qian-Shun Dai}
\author[a,b,c,1]{and Ding Yu Shao\note{Corresponding author.}}

\affiliation[a]{Department of Physics and Center for Field Theory and Particle Physics, Fudan University, Shanghai, China}
\affiliation[b]{Key Laboratory of Nuclear Physics and Ion-beam Application (MOE), Fudan University, Shanghai, China}
\affiliation[c]{Shanghai Qi Zhi Institute, Shanghai 200030, China}

\emailAdd{chengzhang\_phy@fudan.edu.cn, qsdai21@m.fudan.edu.cn, dingyu.shao@cern.ch}

\abstract
{We study the azimuthal angular decorrelation of the dijet production via photon fusion in ultra-peripheral heavy ion collisions. The impact parameter dependent cross section of quark-antiquark pairs production is derived using the equivalent photon approximation, and the contribution from final-state QCD radiations to the azimuthal angular distribution are calculated within Soft-Collinear Effective Theory. We carry out the QCD resummation of large logarithms of the azimuthal angle as well as the jet radius at next-to-leading logarithmic accuracy. In the end we present the normalized differential cross section for azimuthal decorrelation of the dijet pair and find that our results are consistent with the measurements reported by the ATLAS collaboration.}

\begin{document}
\maketitle

\section{Introduction}

\label{sec:intro}

Ultraperipheral collisions (UPCs) of high-energy nuclei are processes
that the impact parameter is large enough so that only electromagnetic
interactions happen with hadronic interactions being excluded. UPCs provide
unique opportunities to study the strong electromagnetic
processes in relativistic heavy-ion collisions \cite{Baur:1998ay,Baur:2003ar,Bertulani:2005ru,Baltz:2007kq,Zhao:2022dac}. Besides, UPCs
involve two-photon fusions and photonuclear interactions, which provide
an opportunity to study the quasi-real photons emitted by one of the
colliding nuclei and probe the nuclear parton distribution functions
(PDFs). Typical photonuclear processes~\cite{ZEUS:2002wfj,PHENIX:2009xtn,Contreras:2013oan,Schmidke:2016ccw,CMS:2016itn,ALICE:2019tqa,ALICE:2021tyx,LHCb:2021bfl,STAR:2021wwq,STAR:2022wfe}
involve the photoproduction of vector mesons via photon-gluon fusions,
which is a clean probe to study the gluon distributions inside the
energetic nucleus~\cite{Brodsky:1994kf,Klein:1999qj,Kowalski:2006hc,Rebyakova:2011vf,Lappi:2013am,Guzey:2013qza,Xie:2016ino,Xing:2020hwh,Zha:2020cst,Brandenburg:2022jgr,Mantysaari:2022sux}.
Typical two-photon interactions involve dilepton productions~\cite{ATLAS:2022cbd,STAR:2004bzo,STAR:2018ldd,STAR:2018xaj,ATLAS:2018pfw,ALICE:2018ael,ATLAS:2020epq}
or diphoton productions from light-by-light scatterings~\cite{ATLAS:2017fur}.
The dilepton production process can serve to probe the strong electromagnetic
fields generated from the initial state relativistic heavy-ion~\cite{Vidovic:1992ik,Hencken:1994my,Krauss:1997vr,Zha:2018tlq,Klein:2018fmp,Brandenburg:2021lnj,Wang:2021kxm},
as well as the polarizations of these quasi-real photons~\cite{Li:2019sin,Li:2019yzy,Wang:2022gkd},
or to study the final-state Quantum Electrodynamics (QED) radiations
from the produced charged leptons. 

The photon flux coherently emitted from the charged nucleus is highly
enhanced by a factor of the squared nuclear charge $Z^{2}$, and
these photons are nearly real with a small virtuality $k^{2}\lesssim(\hbar/R_{A})^{2}$,
where $R_{A}$ is the nuclear radius. The quasi-real photon flux can
be well described by the well-known equivalent photon approximation
(EPA) introduced by Fermi, von Weizs$\ddot{\textrm{a}}$cker and Williams~\cite{Fermi:1924tc,vonWeizsacker:1934nji,Williams:1935dka},
where the strong photon flux is treated as a highly Lorentz-contracted
classical electromagnetic field emitted from the fast-moving nucleus.
Within the EPA method, the photon spectrum is the Fourier transform
of the nuclear charge spatial distribution.

Recently, in ultraperipheral $Pb+Pb$ collisions Quantum Chromodynamics (QCD) jets production
in $0n0n$ events was first observed by the ATLAS collaboration~\cite{ATLAS:2022cbd}, and they find \textsc{Pythia}8 Monte Carlo event generator \cite{Bierlich:2022pfr} underestimates the total cross section by about an order of magnitude. Usually,
dijet and multi-jet production in UPCs with no nuclear breakup involves
both photon fusion and diffractive photo-production processes. This
discrepancy in magnitude may result from the missing processes of
the diffractive photo-production in \textsc{Pythia}8, and these
processes has been calculated in \cite{Guzey:2016tek}, where the
results are obtained by using the next-to-leading order (NLO) collinear factorization formalism of
perturbative QCD (pQCD). Besides, in \cite{ATLAS:2022cbd} the ATLAS
collaboration also present the experimental results of the azimuthal
angle $\Delta\phi_{jj}$ between the two jets with the highest transverse momenta. In order to directly
compare the shape of the measured distribution with the \textsc{Pythia} results,
they have rescaled the \textsc{Pythia} curve to match the number of measured
events. As a result, they find the measured distribution is obviously
wider than that from the \textsc{Pythia}8 event generator. Therefore, it is
significant to have a theoretical calculation for the $\Delta\phi_{jj}$-distribution.

Starting with the pioneering papers \cite{Banfi:2008qs,Hautmann:2008vd}, all-order resummation of azimuthal decorrelation for QCD jets has
been performed in various processes \cite{Sun:2014gfa,Sun:2015doa,Chen:2018fqu,Sun:2018icb,Liu:2018trl,Chien:2019gyf,Liu:2020dct,Chien:2020hzh,Abdulhamid:2021xtt,Chien:2022wiq,Bouaziz:2022tik,Yang:2022qgk,Martinez:2022dux}.
Usually, fixed-order calculations in pQCD can be used
to systematically improve the description of hadronic radiation. However,
in nearly back-to-back region $\Delta\phi_{jj}\sim\pi$ the fixed-order
expansion of pQCD diverges due to the existence of logarithms $\alpha_s^n\ln^m(\pi-\Delta\phi_{jj})$ (with $0<m \leq 2n$) at each perturbative order $n$, so that an all-order resummation is necessary
for the validity of theoretical predictions. Besides, as pointed in \cite{Banfi:2008qs,Chien:2022wiq}
the all-order resummation formula strongly depends on the recombination
schemes in sequential jet clustering algorithms. Especially when the
jets are reconstructed with the $p_{T}^n$-weighted recombination scheme,
the resummation is straightforward
since it does not require any non-trivial treatment of non-global
logarithms (NGLs) \cite{Dasgupta:2001sh,Banfi:2002hw}.

\begin{figure}[t]
\centering \includegraphics[clip,width=0.85\textwidth]{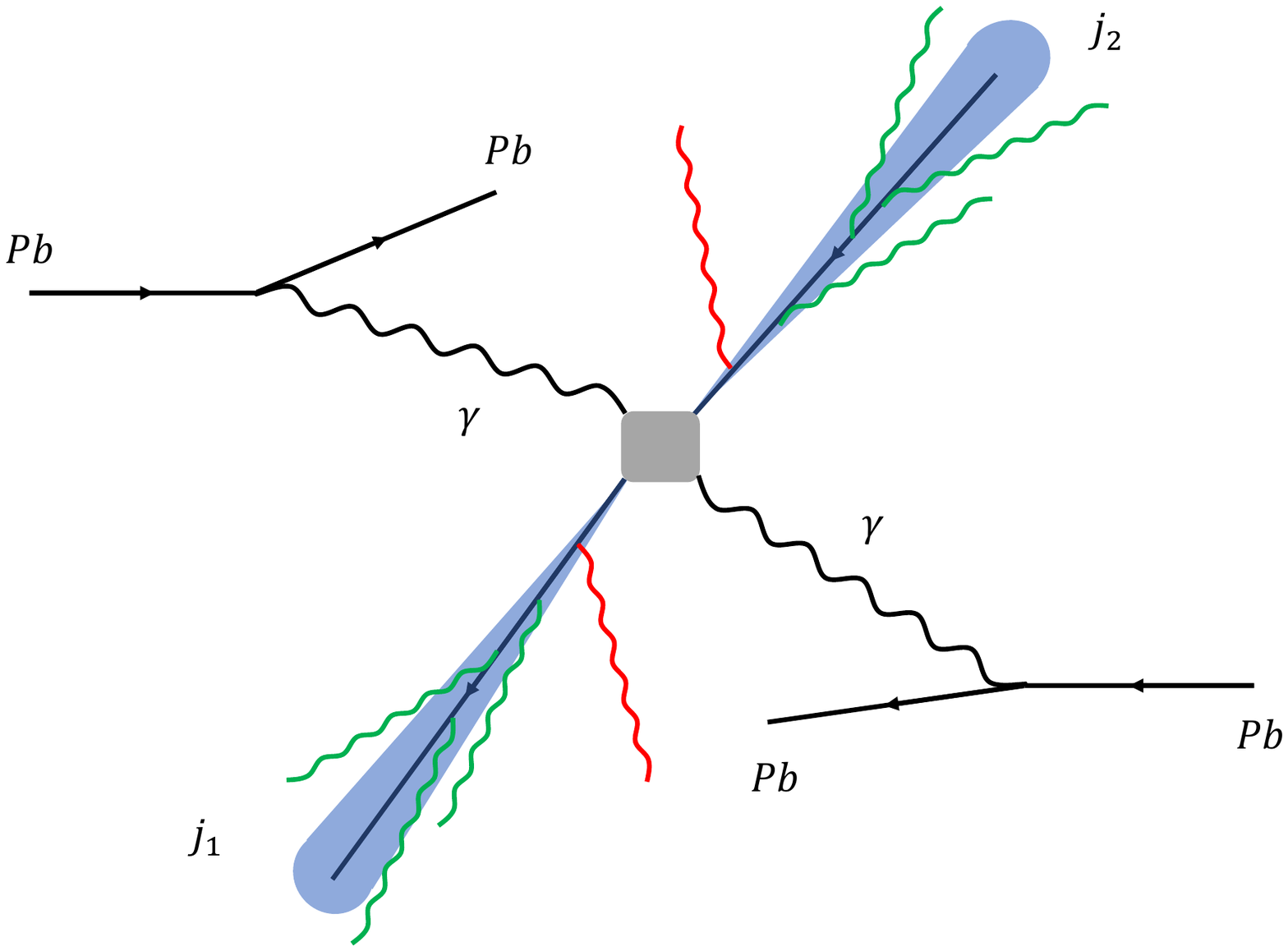}
\vspace{-1.5cm}
\caption{Photon-induced dijet production in UPCs.}
\label{fig:dijetsUPC} 
\end{figure}

In this paper we perform a detailed theoretical study for the azimuthal
decorrelation of dijet productions from photon-photon fusion in ultraperipheral
heavy ion collisions, shown in figure \ref{fig:dijetsUPC}. Specifically, we consider the $\Delta\phi_{jj}$-distribution
of the leading dijet pair produced nearly back to back in the transverse
plane to the beam. In order to describe such configuration we need
to consider nonzero transverse momenta from both incoming and outgoing
partons. First of all, since the initial photons can have nonzero transverse
momentum, one needs to consider the transverse momentum dependent
photon distribution. This effect can be captured by the impact parameter
dependent cross section in EPA, and the corresponding calculations
have been extensively studied in \cite{Vidovic:1992ik,Krauss:1997vr,Zha:2018tlq,Li:2019yzy,Li:2019sin,Zha:2020cst,Brandenburg:2021lnj,Wang:2021kxm,Wang:2022gkd}. Moreover, final-state radiations could also introduce azimuthal
angular decorrelation. We apply Soft-Collinear Effective Theory (SCET)~\cite{Bauer:2000yr,Bauer:2001ct,Bauer:2001yt,Bauer:2002nz,Beneke:2002ph}
to develop a factorization and resummation formalism for studying the azimuthal angular
distribution in back-to-back dijet production in ultraperipheral heavy
ion collisions, where large logarithms of the azimuthal angle and the jet radius are resumed at next-to-leading logarithmic (NLL) accuracy. 

The rest of this paper is organized as follows. In section \ref{sec:born}
we discuss the impact parameter dependent cross section of quark-antiquark pairs 
production via photon-photon fusions in UPCs. In section \ref{sec:fac-res}
we give the factorization and resummation formula of $\gamma\gamma\to2$
jets within SCET. We present the numerical results using the theoretical
formula, enumerate all theoretical uncertainties and compare our predictions
with the ATLAS experimental data in section \ref{sec:numerics}. We
conclude in section \ref{sec:conclusion}. The details of the one-loop
calculations, resummation, and anomalous dimensions are provided in
the appendix.

\section{Quark-antiquark pairs production from $\gamma\gamma$ fusions}

\label{sec:born}

\begin{figure}[t]
  \centering \includegraphics[clip,width=0.41\textwidth]{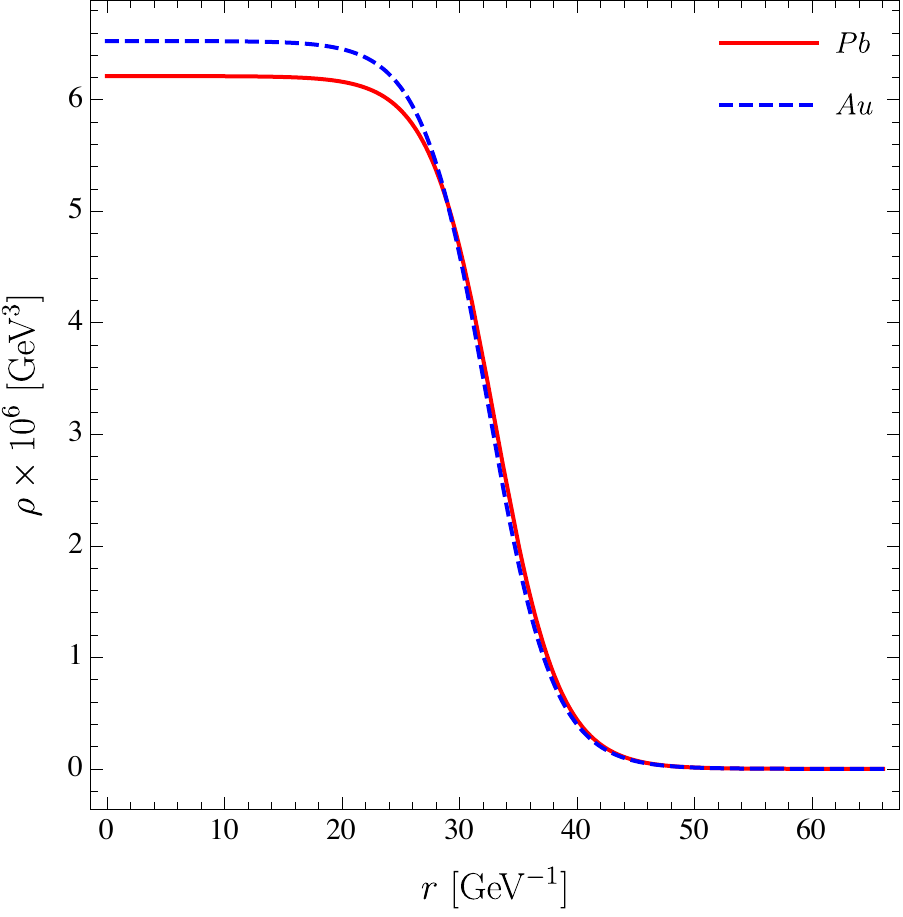}~~~~
  \includegraphics[clip,width=0.42\textwidth]{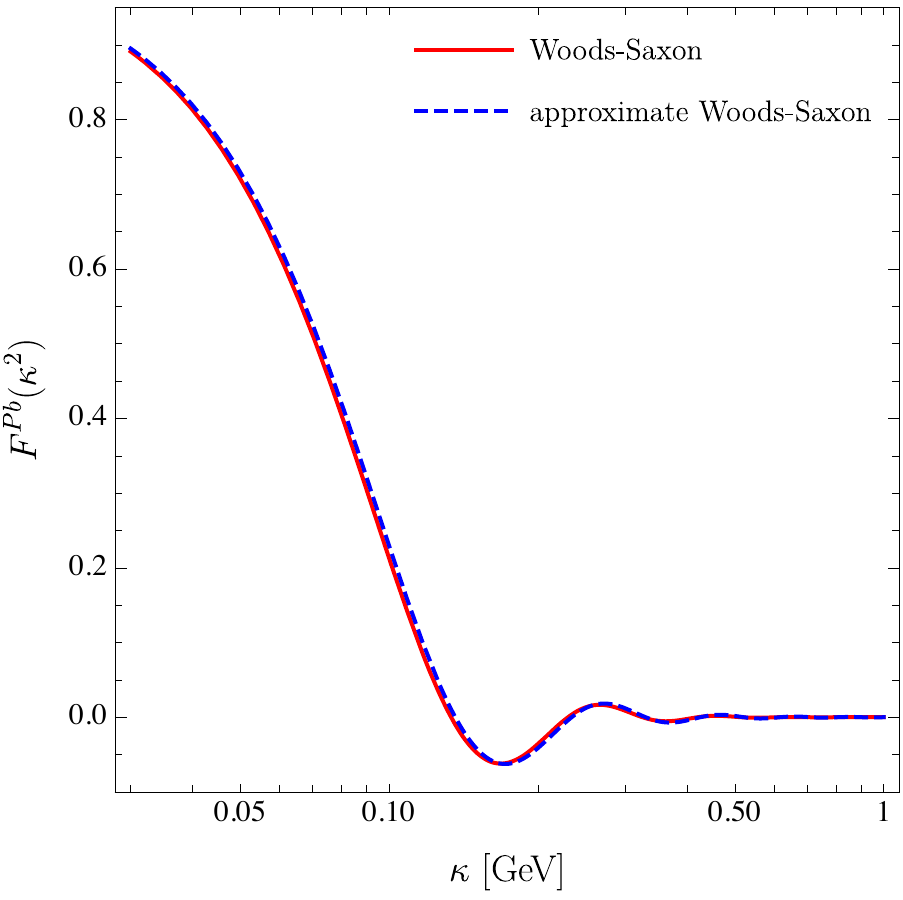} \caption{Left: The nuclear charge spatial distribution $\rho(r)$ within the
  Woods-Saxon model of lead (red solid curve) and gold nuclei
  (blue dashed curve), both of which have been normalized to unity $\int\mathrm{d}^{3}\boldsymbol{r}\rho(r)=1$;
  Right: The lead nuclear charge form factor $F(\kappa^{2})$ of the Woods-Saxon
  type (solid red curve) and from STARlight Monte Carlo event generator
  (dashed blue curve).}
  \label{fig:rho=000026F}
  \end{figure}

In this section, we give the impact parameter dependent cross section
of quark-antiquark pairs production from photon-photon fusions in UPCs at the
lowest QED order, which provides the initial state of the dijet production. Associating that with the final-state QCD emissions, we arrive at
the dijet production from two-photon interactions in UPCs. The two-photon
fusion process reads, 
\begin{equation}
\gamma(\omega_{1},\bm{k}_{1T})+\gamma(\omega_{2},\bm{k}_{2T})\rightarrow q(y_{1},\bm{p}_{1T})+\bar{q}(y_{2},\bm{p}_{2T}).
\end{equation}
After applying the Weizs\"{a}cker-Williams equivalent photon approximation (EPA) \cite{Fermi:1924tc,vonWeizsacker:1934nji,Williams:1935dka}, one can drive the unpolarized UPC differential cross section as~\cite{Krauss:1997vr,Li:2019yzy,Li:2019sin} 
\begin{align}\label{eq:b-dependent-xsetion}
\frac{\mathrm{d}^{5}\sigma_{0}}{\mathrm{d}^{2}\bm{q}_{T}\mathrm{d}p_{T}\mathrm{d}y_{1}\mathrm{d}y_{2}}= & \,N_{c}\sum_{q}e_{q}^{4}\frac{Z^{4}\alpha_{{\rm em}}^{4}}{\pi^{5}M^{4}}p_{T}\int \mathrm{d}^{2}\bm{b}_{T}\mathrm{d}^{2}\bm{k}_{1T}\mathrm{d}^{2}\bm{k}_{2T}\mathrm{d}^{2}\bm{k}_{1T}'\mathrm{d}^{2}\bm{k}_{2T}'\nonumber \\
 & \times\delta^{(2)}(\bm{k}_{1T}+\bm{k}_{2T}-\bm{q}_{T})\,\delta^{(2)}(\bm{k}_{1T}'+\bm{k}_{2T}'-\bm{q}_{T})\,e^{i\,(\bm{k}_{1T}-\bm{k}_{1T}')\cdot\bm{b}_{T}}\nonumber \\
 & \times k_{1T}\frac{F(-k_{1}^{2})}{-k_{1}^{2}}k_{2T}\frac{F(-k_{2}^{2})}{-k_{2}^{2}}k_{1T}'\frac{F(-k_{1}'^{2})}{-k_{1}'^{2}}k_{2T}'\frac{F(-k_{2}'^{2})}{-k_{2}'^{2}}\frac{M^{2}-2p_{T}^{2}}{p_{T}^{2}}\nonumber \\
 & \times\cos\left(\phi_{\bm{k}_{1T}}-\phi_{\bm{k}_{1T}'}+\phi_{\bm{k}_{2T}}-\phi_{\bm{k}_{2T}'}\right),
\end{align}
with the nuclear electric charge $Z$ ($79$
for $Au$, $82$ for $Pb$), the fine-structure constant $\alpha_{{\rm em}}=1/137$,
the number of colors $N_{c}=3$, the quark fractional electric charge
$e_{q}$, and the quark flavors $q$. Since we only consider QCD light flavor jet production, the cross section is a sum over the light quark flavors $q=u, d, s$. Here we
define the average transverse momentum $\bm p_T$ of the two jets and the transverse momentum imbalance $\bm q_T$ as follows
\begin{align}
  \boldsymbol{p}_{T}\equiv\left(\boldsymbol{p}_{1T}-\boldsymbol{p}_{2T}\right)/2,~~~
  \boldsymbol{q}_{T}\equiv\boldsymbol{p}_{1T}+\boldsymbol{p}_{2T}.
\end{align}
The quark-antiquark pairs are produced nearly back-to-back in the transverse plane with respect to the beam direction, so the transverse momentum imbalance $q_T\equiv |\bm{q}_T|$ is small. In the limit of $q_T\ll p_T \sim p_{1T} \sim p_{2T}$, the invariant mass of quark-antiquark pairs reads
\begin{align}\label{eq:invM}
  M=p_{T}\sqrt{2+2\cosh(y_{1}-y_{2})},
\end{align}
with quark rapidities $y_{1,2}$. The impact parameter $b_{T}$ should be integrated over from $2R_{A}$
to $\infty$ to exclude the central collisions, where $R_{A}$ is
the nuclear radius. The photon possesses different transverse momentum
$\boldsymbol{k}_{T}$ (with the azimuthal angle $\phi_{\boldsymbol{k}_{T}}$)
and $\boldsymbol{k}_{T}'$ inside the amplitude and the conjugate
amplitude respectively, as we have kept the impact parameter dependence
of two colliding photon fluxes. The quasi-real photon flux with
a small virtuality $k^{2}=-\omega^{2}/\gamma^{2}-\bm k_{T}^{2}$
(with the Lorentz boost factor $\gamma=2676$ at a centre-of-mass
energy per nucleon pair of $\sqrt{s_{\mathrm{NN}}}=5.02$ TeV at the
CERN LHC) and photon energy $\omega_{1,2}=p_{T}(e^{\pm y_{1}}+e^{\pm y_{2}})/2$
can be described by the EPA, which depicts the photon momentum $k$ distribution via the
nuclear charge form factor $F(-k^{2})$. This form factor can be obtained after performing Fourier transformation from the nuclear charge distribution $\rho(r)$ as 
\begin{equation}
F(\bm k^{2})=\int\mathrm{d}^{3}\boldsymbol{r}e^{i\boldsymbol{r}\cdot\boldsymbol{k}}\rho(r),
\end{equation}
where the charge distribution inside the heavy nucleus is well described
by the Woods-Saxon distribution, 
\begin{equation}
\rho_{\mathrm{WS}}(r)=\frac{1}{1+e^{(r-R_{A})/d}}\Bigg / \!\! \int\mathrm{d}^{3}\boldsymbol{r}\frac{1}{1+e^{(r-R_{A})/d}},
\end{equation}
with the nuclear radius $R_{A}\simeq 1.2\,A^{1/3}$ (where $A$ is the nucleon number) and the nuclear skin depth $d$ ($d=0.546$ fm for $Pb$ and $d=0.535$ fm for $Au$). In the left panel of figure \ref{fig:rho=000026F} we plot the nuclear charge spatial distribution of lead (red solid curve) and gold (blue dashed curve) nuclei within the Woods-Saxon model. One can see that the lead nuclei possesses a slightly wider spatial broadening and a less steep boundary compared to the gold nuclei. For simplicity, in the latter calculation, we alternatively choose the approximate Woods-Saxon form factor used in the STARlight Monte Carlo event generator~\cite{Klein:2016yzr},
\begin{equation}
F_{\rm AWS}(\kappa^{2})=\frac{3[\sin(\kappa R_{A})-\kappa R_{A}\cos(\kappa R_{A})]}{(\kappa R_{A})^{3}(a^{2}\kappa^{2}+1)},
\end{equation}
with $a=0.7$ fm. The corresponding nuclear charge form factors are shown in the right panel of figure \ref{fig:rho=000026F}, and we find the difference between these two models is negligible. From the plot we find that the typical transverse momentum of the photon is about $50$ MeV for $Pb$, and then the corresponding azimuthal angular is too small and can not explain the results observed by the ATLAS collaboration. Therefore we also need to systematically calculate higher-order QCD and QED corrections in order to describe data at the LHC. In the following section, we will consider the all-order resummation corrections from final-state QCD radiations, which play a dominant role in jet production, and we leave a full QCD and QED showers study for the future work.

\section{Factorization and resummation for final-state radiations}\label{sec:fac-res}

In this section we will present the resummation formula describing final-state QCD emissions from the dijet system, which contributes to the azimuthal decorrelation of the dijet pair. In order to obtain such formalism, we first apply SCET to write down a factorized formula, and express the differential cross section as the product of single-scale key ingredients at leading power. Based on the factorization formula and requirements of the factorization scale independence for the cross section, we obtain the all-order resummation formula via solving corresponding renormalization group (RG) equations. 

Before introducing our methodology, we first briefly describe two methods on the market used to perform resummation of large logarithms of the azimuthal angle for jets \cite{Banfi:2008qs,Hautmann:2008vd,Sun:2014gfa,Sun:2015doa,Chen:2018fqu,Sun:2018icb,Liu:2018trl,Chien:2019gyf,Liu:2020dct,Chien:2020hzh,Abdulhamid:2021xtt,Chien:2022wiq,Bouaziz:2022tik,Yang:2022qgk,Martinez:2022dux}. The first method is deriving an all-order resummation formula for the transverse momentum imbalance $\bm q_T$, and then constructing the azimuthal decorrelation $\Delta\phi$ distribution from the $\bm q_T$ distribution, which can be named as the ``indirect method". The second method is calculating the resumed results of the azimuthal angular distribution directly, so we name it as the ``direct method". The relation between the two methods is obvious for Drell-Yan-like processes, but for the scattering processes involving jet production it is not easy to establish the relation, since one also needs to carry out the resummation of large logarithms from final-state QCD radiations. As shown in \cite{Buffing:2018ggv,Chien:2019gyf,delCastillo:2020omr}, in the indirect method the single logarithmic anomalous dimensions of the soft and collinear-soft function with the narrow cone approximation both depend on the azimuthal angle of $\bm q_T$ (or the azimuthal angle of $\bm b$ in the Fourier conjugate transverse coordinate space). After taking into account QCD evolution, this azimuthal integral is divergent in some phase space regions. In order to regularize such divergences, different schemes have been proposed in \cite{Buffing:2018ggv,Chien:2019gyf,delCastillo:2021znl}. In this paper we consider QCD resummation of large logarithms of azimuthal angle and jet radius at the NLL accuracy. Therefore, in order to avoid such difficulties, we will apply the direct method in the latter calculations. 

\subsection{QCD factorization formalism in SCET}

A factorization formula for the boson-jet azimuthal decorrelation has been comprehensively derived by one of us in \cite{Chien:2022wiq} within SCET, for the case where jets are defined using the anti-$k_T$ algorithm \cite{Cacciari:2008gp} with the $p_T^n$-weighted recombination scheme. Although in this paper we consider the jet definition using the standard recombination schemes, which is different from $p_T^n$-weighted scheme, the factorization formula still shares many similar properties. Therefore, we only present the main features of the factorized structure, and a more detailed discussion on the factorization analysis within SCET can be found in \cite{Chien:2022wiq}.

At the first step, we use scaling arguments to identify the regions of phase space that contribute to the factorization formula at the leading power. In the back-to-back limit where $\Delta \phi_{jj} \to \pi$, the relevant low energy modes for the factorized expression in SCET are given by 
\begin{align}
  { \color{RoyalBlue} n_i \textbf{ collinear}}:&~~p_{c_i}^\mu\sim p_T\,(R^2,1,R)_{n_i \bar n_i},  \label{eq:coll-mode}\\
  { \color{ForestGreen} n_i \textbf{ collinear-soft}}:&~~ p_{cs_i}^\mu\sim \frac{p_T\,\delta\phi}{R}(R^2,1,R)_{n_i \bar n_i}, \label{eq:coft-mode} \\
  {\color{red} \textbf{soft}}:&~~p_s^\mu\sim p_T\,(\delta\phi,\delta\phi,\delta\phi), \label{eq:soft-mode}
\end{align}
with $\delta\phi \equiv\pi-\Delta\phi_{jj}$. Here all the momenta $p^\mu \equiv (n_i\cdot p,\bar n_i\cdot p, p_{n_i\perp})_{n_i \bar n_i}$ are expressed using light-cone coordinates, where $n_1^\mu$ and $n_2^\mu$ are light-like vectors along leading and sub-leading jets, separately, and $\bar n_i^\mu$ is the direction backwards to the jet.  In the above analysis we have considered the limits $R\ll1$, therefore the narrow cone approximation as well. As a result, the $n_i$ collinear mode \eqref{eq:coll-mode} describes energetic emissions inside the jet with radius $R$, which only contributes to the normalization of the $\Delta\phi_{jj}$-distribution, not to its shape. While both $n_i$ collinear-soft and soft modes contribute to $\Delta\phi_{jj}$-distribution, where $p_{cs_i}^\mu$ is sensitive to the jet boundary but $p_s^\mu$ is not. Besides, the soft mode describes large-angle radiations without any direction preference, so in \eqref{eq:soft-mode} the subscript of $p_s^\mu$ in light-cone coordinates can be ignored.

With the relevant scaling identified, we can write down a factorized form for the differential cross section at the leading power of $\delta\phi$
\begin{align}\label{eq:fac_mom}
  \frac{\mathrm{d}^4\sigma}{\mathrm{d} q_x \mathrm{d} p_T \mathrm{d} y_1 \mathrm{d} y_2} = \,& \int \mathrm{d}k_x \, \mathrm{d}\lambda_x \, \mathrm{d}l_{1,x} \, \mathrm{d}l_{2,x} \delta(q_x+\lambda_x+l_{1,x}+l_{2,x}-k_x) B(k_x,p_T,y_1,y_2) \notag \\
  & \times H(p_T,\Delta y,\mu) S(\lambda_x,y_1,y_2,\mu,\nu) U_1(l_{1,x},R,y_1,\mu,\nu) J_1(p_T,R,\mu) \notag \\
  & \times U_2(l_{2,x},R,y_2,\mu,\nu) J_2(p_T,R,\mu),
\end{align}
with $|q_x| \approx p_T \sin(\pi-\Delta \phi_{jj})=p_T\sin\delta\phi$ as shown in the left panel of figure \ref{fig:def}, where we take the leading jet's transverse momentum as reference for $-y$ direction,
\begin{align}
  \bm p_{1T} \approx (0, - p_T),~~~ \bm p_{2T} \approx p_T ( \sin\delta\phi, \cos\delta\phi). 
\end{align}
In \eqref{eq:fac_mom} the Dirac delta function term in the first line indicates the momentum conservation of different ingredients along the $x$-axis, neglecting power corrections. Explicitly, the pieces of the factorized formula \eqref{eq:fac_mom} are:

\begin{figure}[t]
  \centering
  \vspace{-2.5cm}
  \includegraphics[width=1.1\textwidth,clip]{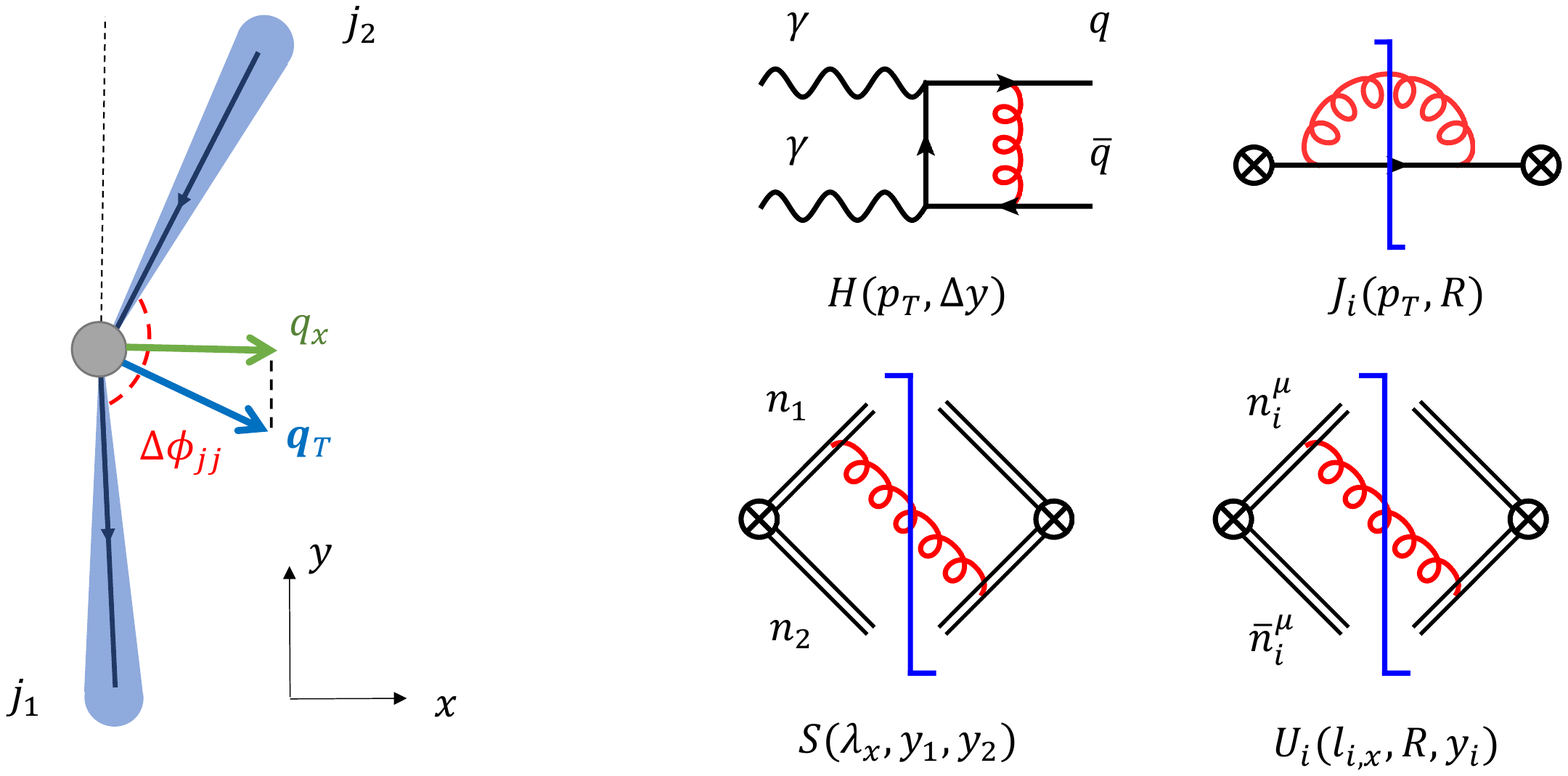}
  \vspace{-3cm}
  \caption{Left: The definition of the azimuthal angle $\Delta\phi_{jj}$ between leading $j_1$ and subleading jet $j_2$ in $x$-$y$ plane, and the relation between $\Delta\phi_{jj}$ and momentum $q_x$ in the factorized formula \eqref{eq:fac_mom}, where $\bm q_T$ is the vector sum of the transverse momenta for the dijet, and $q_x$ is the projection of $\bm q_T$ in the $x$-axis.  Right: Sample one-loop diagrams of the hard, jet, soft as well as collinear-soft functions in \eqref{eq:fac_mom}. }\label{fig:def}
\end{figure}

\begin{itemize}
  \item $B(k_x,p_T,y_1,y_2)$ is the Born cross section of the process $\gamma\gamma \to q \bar q$ calculated in section \ref{sec:born}, and $k_x$ indicates $x$ component of the transverse momentum from incoming photon beams. Explicitly, we obtain the function $B(k_x,p_T,y_1,y_2)$ from \eqref{eq:b-dependent-xsetion} as
  \begin{align}
    B(k_x,p_T,y_1,y_2) = \int \mathrm{d} k_y \frac{\mathrm{d}^{5}\sigma_{0}}{\mathrm{d}^{2}\bm{k}_{T}\mathrm{d}p_{T}\mathrm{d}y_{1}\mathrm{d}y_{2}},~~~{\rm with}~~~ \bm k_T=(k_x,k_y).
  \end{align}
  \item $H(p_T,\Delta y)$ is the hard function that can be determined order-by-order in perturbation theory by a matching calculation in QCD and in SCET at the hard scale $\mu_h$. Here $\Delta y\equiv y_1-y_2$, is defined as the rapidity difference between leading and sub-leading jets. 
  \item $J_i(p_T,R)$ is the jet function describing the emission of collinear radiations inside the anti-$k_T$ jet with radius $R$, and its mode has momenta that scale as $p_{c_i}^\mu$. 
  \item $U_i(l_{i,x},R,y_i)$ is the collinear-soft function describing the soft radiation along the direction of the jet, and it is sensitive to the jet direction and the jet boundary. In addition, the collinear-soft mode $p_{cs_i}^\mu$ can also resolve any possible collinear constituents of the jet, which can give the so-called NGLs, as discussed in the following paragraph.    
  \item $S(\lambda_x,y_1,y_2)$ is the transverse momentum dependent soft function that integrates the radiation from the leading and sub-leading jets, so it depends on the rapidity of jets explicitly. 
\end{itemize}
Their one-loop sample diagrams in SCET are given in the right panel of figure \ref{fig:def}, and in the appendix \ref{app:one-loop-funs} we present the explicit calculations of the soft and collinear-soft functions at one loop. In addition to the factorization scale $\mu$ dependence, we also present the soft $S$ and collinear-soft $U_i$ functions with explicit rapidity scale $\nu$ dependence, which stems from rapidity divergence and the corresponding regulator. In the following subsection we will apply the collinear anomaly formalism in \cite{Becher:2010tm,Becher:2011xn} to resum corresponding rapidity logarithms. Alternatively, it can be dealt with using the rapidity renormalization group method \cite{Chiu:2011qc,Chiu:2012ir}. Finally, it should be noted that the above factorized expression \eqref{eq:fac_mom} has ignored the structure from non-global logarithms (NGLs), which start contributing at two-loop order \cite{Dasgupta:2001sh}. The TMD factorization formula including those effects has been discussed in \cite{Liu:2018trl,Chien:2019gyf,Kang:2020xez}, and one finds that NGLs can be resumed via a fitting function given in \cite{Dasgupta:2001sh} at NLL level. In our phenomenology, we have included their contributions in the resummation formula.

After performing Fourier transform for \eqref{eq:fac_mom}, we obtain the factorized formula in the coordinate space as follows
\begin{align}\label{eq:fac}
  \frac{\mathrm{d}^4\sigma}{\mathrm{d} q_x \mathrm{d} p_T \mathrm{d} y_1 \mathrm{d} y_2} =&  \int_{-\infty}^{+\infty} \frac{\mathrm{d}b_x}{2\pi}e^{i q_x b_x} \tilde B(b_x,p_T,y_1,y_2) H(p_T,\Delta y,\mu) \tilde S(b_x,y_1,y_2,\mu,\nu) \notag \\
  & \times  \tilde U_1(b_x,R,y_1,\mu,\nu) J_1(p_T,R,\mu) \tilde U_2(b_x,R,y_2,\mu,\nu) J_2(p_T,R,\mu),
\end{align}
where $\tilde B$, $\tilde S$ and $\tilde U_i$ are the Fourier transform of $B$, $S$ and $U_i$ in \eqref{eq:fac_mom}, respectively. Except for the Born cross section $\tilde B$, all other ingredients are normalized to $1$ at the leading order. Accordingly, as a check one can easily see that at the leading order the above formula \eqref{eq:fac} degenerate \eqref{eq:b-dependent-xsetion} for the production of the quark-antiquark pair without final-state radiations.

\subsection{QCD resummation formalism of final-state radiation}

In this section, we present the RG equations for the factorization scale dependent ingredients in \eqref{eq:fac}, including the hard function $H$, the jet function $J_i$, the soft function $\tilde S$, and collinear-soft function $\tilde U_i$. After presenting their RG evolution equations, we check the RG consistency at one loop. In the end, we present the all-order QCD resummation formula for the azimuthal angular distribution.

For the hard scattering process $\gamma\gamma\to q\bar q$, the QCD corrections have been perturbatively calculated at three loop \cite{Caola:2020dfu}, and the corresponding RG equation of the hard function reads
\begin{align}\label{eq:hard-RG}
  \frac{\mathrm{d}}{\mathrm{d\ln\mu}} H(p_T,\Delta y,\mu)  = \underbrace{\left[ -2 C_F \gamma_{\rm cusp}(\alpha_s) \ln \frac{\mu^2}{M^2} + 4 \gamma_q(\alpha_s) \right]}_{\let\scriptstyle\textstyle\substack{\equiv\Gamma_H(\alpha_s)}} H(p_T,\Delta y,\mu),
\end{align}
where we have defined the hard anomalous dimension $\Gamma_H(\alpha_s)$, and $M$ is the invariant mass of dijet pair defined in \eqref{eq:invM}. In the appendix \ref{app:anoma_dim} we present the perturbative expression of all relevant anomalous dimensions for the NLL resummation. 

The one-loop quark jet function for the anti-$k_T$ algorithm with radius $R$ is calculated in \cite{Ellis:2010rwa}, and it satisfies the RG evolution equations
\begin{align}\label{eq:jet-RG}
  \frac{\mathrm{d}}{\mathrm{d\ln\mu}} J_i(p_T,R,\mu) & = \underbrace{\left[ - C_F \gamma_{\rm cusp}(\alpha_s) \ln \frac{p_T^2R^2}{\mu^2} - 2 \gamma_q(\alpha_s) \right]}_{\let\scriptstyle\textstyle\substack{\equiv\Gamma_J(\alpha_s)}} J_i(p_T,R,\mu). 
\end{align}
From \eqref{eq:hard-RG} and \eqref{eq:jet-RG} one can see that the characteristic scales related to the hard and jet functions would be $\mu_h\sim M$ and $\mu_j\sim p_T R$, respectively.  

Due to rapidity divergences, the calculation of soft and collinear-soft functions involves extra complication which is not seen in the calculation of the hard and jet functions. In appendix \ref{app:one-loop-funs} we present their one-loop calculations in \eqref{eq:softnlo} and \eqref{eq:coftnlo}, where we introduce a rapidity regulator by modifying the phase-space integrals \cite{Bell:2018oqa}.
\begin{align}\label{eq:rap-reg}
  \int \mathrm{d}^d k \to \int \mathrm{d}^d k \left(\frac{\nu}{2k^0}\right)^\eta. 
\end{align}
Such divergences are artificial because the product of the soft and collinear-soft functions is independent on the scale $\nu$. However, the rapidity divergence introduces additional logarithmic dependence of the jet radius $R$, so \eqref{eq:fac} does not achieve complete factorization. By refactorizing out $R$-dependence terms within the collinear anomaly framework \cite{Becher:2010tm,Becher:2011xn}, the product of soft and collinear-soft functions should be expressed as
\begin{align}\label{eq:collinear_anomaly}
  \tilde U_1(b,R,y_1,\mu,\nu) \tilde U_2(b,R,y_2,\mu,\nu) \tilde S(b,y_1,y_2,\mu,\nu) = R^{2F_{q\bar q}(b,\,\mu)} W(b,\Delta y,\mu),
\end{align}
where the reminder function $W(b,\Delta y,\mu)$ no longer contains large logarithms of the jet radius, and the relevant $R$-dependent logarithms are resummed by the help of the anomaly exponent $F_{q\bar q}(b,\mu)$. Explicitly, their one-loop expressions are given by
\begin{align}
  F_{q\bar q}(b,\mu) & = \frac{\alpha_s}{4\pi}C_F \gamma_0^{\rm cusp} \ln\frac{b^2\mu^2}{b_0^2} + \mathcal{O}(\alpha_s^2), \label{eq:Fqq} \\
  W(b,\Delta y,\mu) & = 1 - \frac{\alpha_s}{4\pi}C_F \left[ \gamma_0^{\rm cusp} \ln (2+2 \cosh\Delta y) \ln\frac{b^2\mu^2}{b_0^2} \right] + \mathcal{O}(\alpha_s^2), \label{eq:W-fun}
\end{align}
where the reminder function $W$ is normalized to $1$ at the leading order, and they satisfy the following RG equations:
\begin{align}
  \frac{\mathrm{d}}{\mathrm{d\ln\mu}} F_{q\bar q}(b,\mu) & = 2 C_F \gamma_{\rm cusp}(\alpha_s), \\
  \frac{\mathrm{d}}{\mathrm{d\ln\mu}} W(b,\Delta y,\mu) & = \Big[\!-2 C_F \gamma_{\rm cusp}(\alpha_s) \ln (2+2 \cosh\Delta y)\Big] W(b,\Delta y,\mu).
\end{align}
In appendix \ref{app:one-loop-funs} we present the one-loop calculation of the soft and collinear-soft functions, and one can easily verify that the rapidity poles are canceled in the product of $\tilde S$ and $\tilde U_i$, and also reproduce the refactorization formula \eqref{eq:collinear_anomaly}. 

With the anomalous dimensions presented for all the ingredients, one can easily verify that the factorized formula given in \eqref{eq:fac} satisfies the consistency relations for the RG evolution
\begin{align}
  \frac{\mathrm d}{\mathrm{d} \ln \mu} \left[ R^{2F_{q\bar q}(b,\,\mu)} W(b,\Delta y,\mu) H(p_T,\Delta y,\mu) J_1(p_T,R,\mu) J_2(p_T,R,\mu) \right]=0.
\end{align}
In the end all large logarithms can be resummed by evolving the hard and jet functions from their intrinsic scales $\mu_h$ and $\mu_j$ to the scale of the reminder function and the anomaly exponent at $\mu_b$ separately, and at NLL accuracy we have
\begin{align}\label{eq:NLLres}
  \frac{\mathrm{d}^4\sigma^{\rm NLL}}{\mathrm{d} q_x \mathrm{d} p_T \mathrm{d} y_1 \mathrm{d} y_2} =& \int_{0}^{\infty} \frac{\mathrm{d}b_x}{\pi}\cos(q_x b_x) \tilde B(b_x,p_T,y_1,y_2) \notag\\
  &\times\exp\left[ \int_{\mu_h}^{\mu_b} \frac{\mathrm{d}\mu}{\mu} \, \Gamma_H(\alpha_s) + 2 \int_{\mu_j}^{\mu_b}\frac{\mathrm{d}\mu}{\mu}\,\Gamma_J(\alpha_s)\right]U_{\rm NG}^2(\mu_b,\mu_j),
\end{align}
where we have incorporated NGLs resummation effects included by the function $U_{\rm NG}$. As shown in \cite{Liu:2018trl,Chien:2019gyf,Kang:2020xez} the fitting function given in \cite{Dasgupta:2001sh} can be used to capture leading-logarithmic NGLs after choosing proper initial and final evolution scales. In our process, the resummation of NGLs comes from a non-linear RG evolution between the jet and the collinear-soft function \cite{Chien:2019gyf}, so we choose the jet scale $\mu_j$ and $\mu_b$ for the reminder function in the function $U_{\rm NG}$ that is given by 
\begin{align}
  U_{\mathrm{NG}}\left(\mu_{b}, \mu_{j}\right)=\exp \left[-C_{A} C_F \frac{\pi^{2}}{3} u^{2} \frac{1+(a u)^{2}}{1+(b u)^{c}}\right],
\end{align}
with $a=0.85 \, C_A$, $b=0.86\,C_A$, $c=1.33$ and $u=\ln[\alpha_s(\mu_b)/\alpha_s(\mu_j)]/\beta_0$. Since there are two jet functions in the factorized formula \eqref{eq:fac}, we need to include the square of $U_{\rm NG}$ to take into account the NGL resummation associated with each jet.

\section{Numerical results}\label{sec:numerics}

In this section we will present the numerical results using the resummation formula \eqref{eq:NLLres}, which captures both non-zero transverse momentum from incoming photons and QCD evolution of final-state radiations at NLL accuracy. 

\begin{figure}[t]
  \centering
  \includegraphics[width=0.44\textwidth,clip]{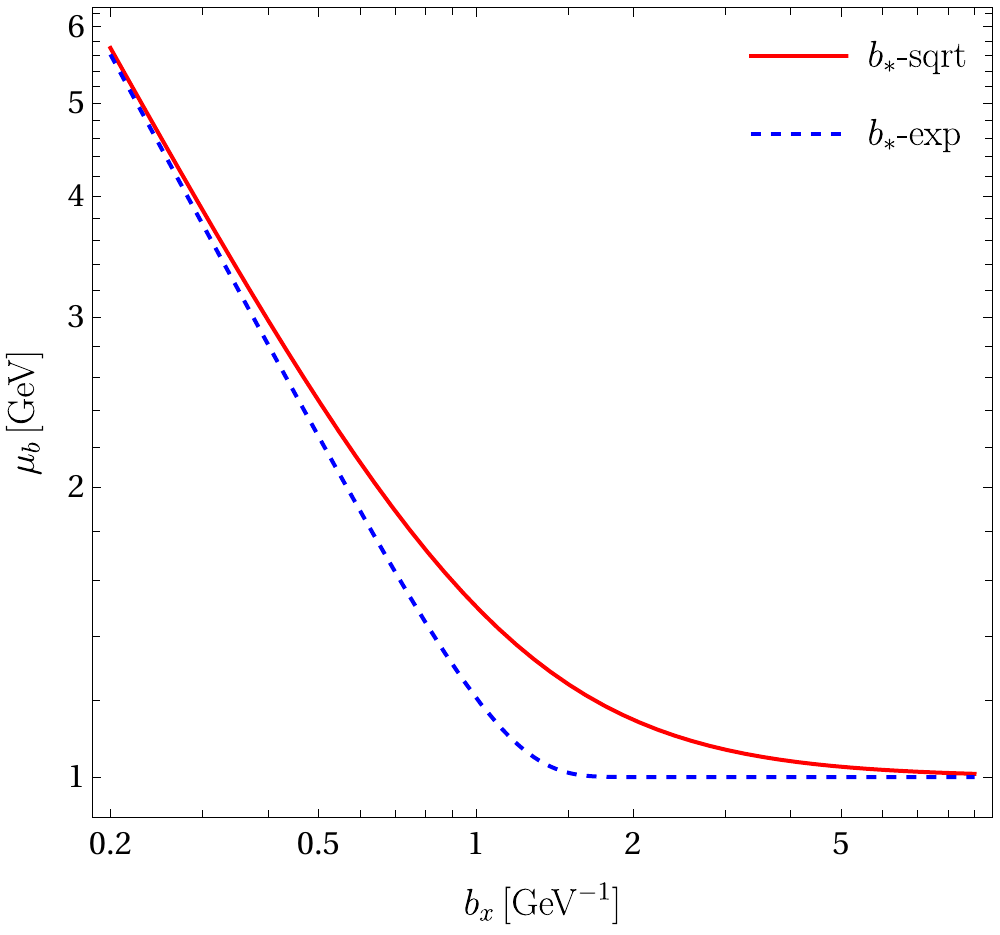}~~~~
  \includegraphics[width=0.45\textwidth,clip]{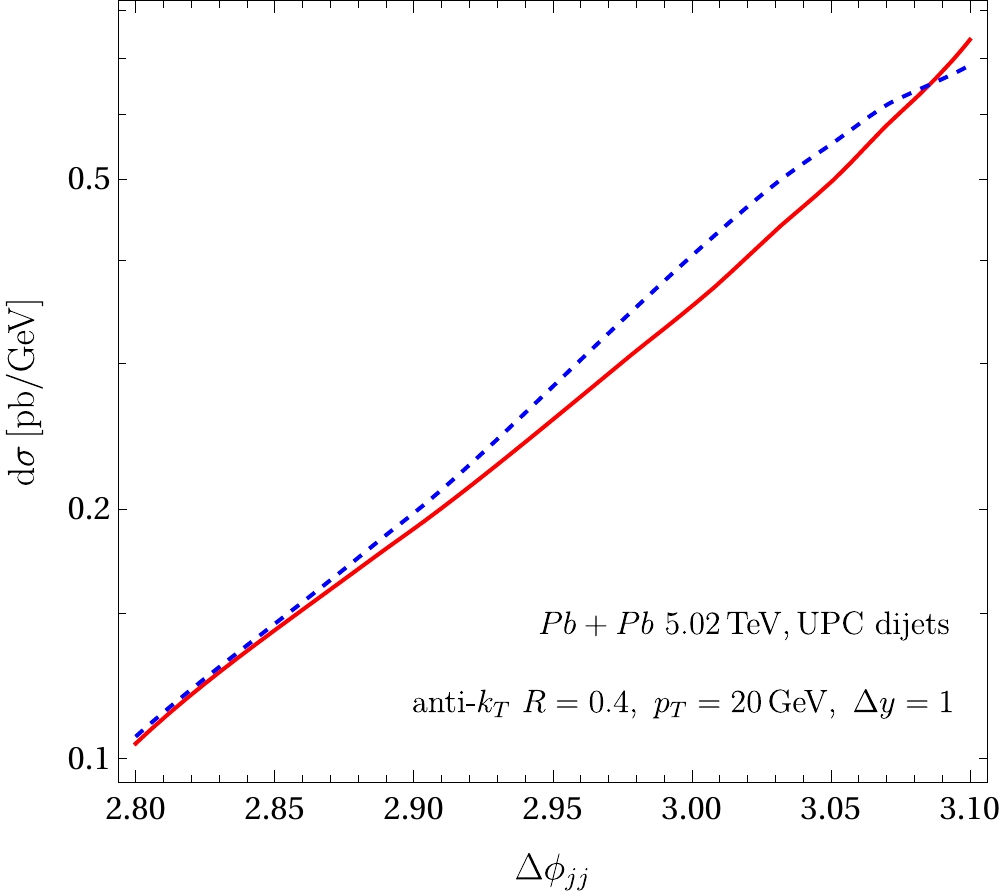}
  \caption{Left: The scale $\mu_b$ in \eqref{eq:scales} depending on the functional form chosen for $b_*$: the red solid line is obtained with the square-root form, while the blue dashed line with the exponential form; see \eqref{eq:squre-root} and \eqref{eq:exp}. Right: Uncertainties of the differential cross section estimated by varying the functional forms in the $b_*$-prescription. In both cases we choose $b_{\rm max}=1.123 \, {\rm GeV}^{-1}$. }\label{fig:mub}
\end{figure}

We see that, after the change of integration variable from $q_x$ to $\Delta\phi_{jj}$ in \eqref{eq:NLLres}, the theoretical results for $\Delta\phi_{jj}$-distribution can be directly obtained where we choose the intrinsic scales in the resummation formula as
\begin{align}\label{eq:scales}
  \mu_h = M,~~~\mu_j=p_T R,~~~\mu_b=\frac{b_0}{b_*(b_x)}.
\end{align}
In the above expressions, we introduce a function $b_*(b_x)$ to freeze the value of $\mu_b$ in the non-perturbative region as $b_x\to \infty$, since pQCD calculations will hit the Landau pole. The standard ``square-root form" \cite{Collins:1984kg} of this function is defined as
\begin{align}\label{eq:squre-root}
    b_*(b) = \frac{b}{\sqrt{1+b^2/b_{\rm max}^2}}.
\end{align}

It is noted that in the original Collins-Soper-Sterman (CSS) resummation, $b_*$-prescription was introduced with additional non-perturbative factors, which are proposed in terms of different functional forms and obtained from fitting experimental data  \cite{Collins:2014jpa,Aidala:2014hva,Sun:2014dqm,Landry:2002ix,Konychev:2005iy,Bacchetta:2017gcc,Bacchetta:2022awv}. However, in our process there are no same double logarithmic terms that from initial-state QCD radiations, so that the nonperturbative contribution would be different from the original ones in the CSS resummation. Therefore, we do not introduce such non-perturbative factors in the numerical results. 

In order to estimate theoretical uncertainties from different function forms of $b_*(b_x)$, we also present the result using the ``exponential form" \cite{Bacchetta:2017gcc}
\begin{align}\label{eq:exp}
  b_*(b) = b_{\rm max} \left[1-\exp\left( - b^4/b_{\rm max}^4 \right)\right]^{1/4}.
\end{align}
to avoid the Landau pole. By introducing the function $b_*(b_x)$, one has $\mu_b \sim b_0/b_x$ when $b_x\ll b_{\rm max}$, and $\mu_b \sim b_0/b_{\rm max}$ when $b_x\gg b_{\rm max}$, as is shown in the left panel of figure \ref{fig:mub}, where the solid red and dashed blue curves are the scales $\mu_b$ with square-root and exponential form, respectively. In both cases, we choose the scale $\mu_b$ freeze at $1$ GeV as $b_x\to \infty$. As expected, these two forms have the same asymptotic behaviors in both large $b_x$ and small $b_x$ regions. The right panel of figure \ref{fig:mub} shows that these two forms lead to similar predictions for the differential cross section in \eqref{eq:NLLres}. From the figure, we find that as $\Delta\phi_{jj} \sim \pi$ the sensitivity of our predictions to this function form is about $10\%$, and when $\Delta\phi_{jj}<2.9$  it has a negligible effect on numerics, where we choose $p_T=20$ GeV and $\Delta y=$1. Since the uncertainty from different function forms is smaller than that from the scale variation, in the rest of this paper we will use the standard ``square-root form" in \eqref{eq:squre-root}.

\begin{figure}[t]
  \centering
  \includegraphics[width=0.65\textwidth,clip]{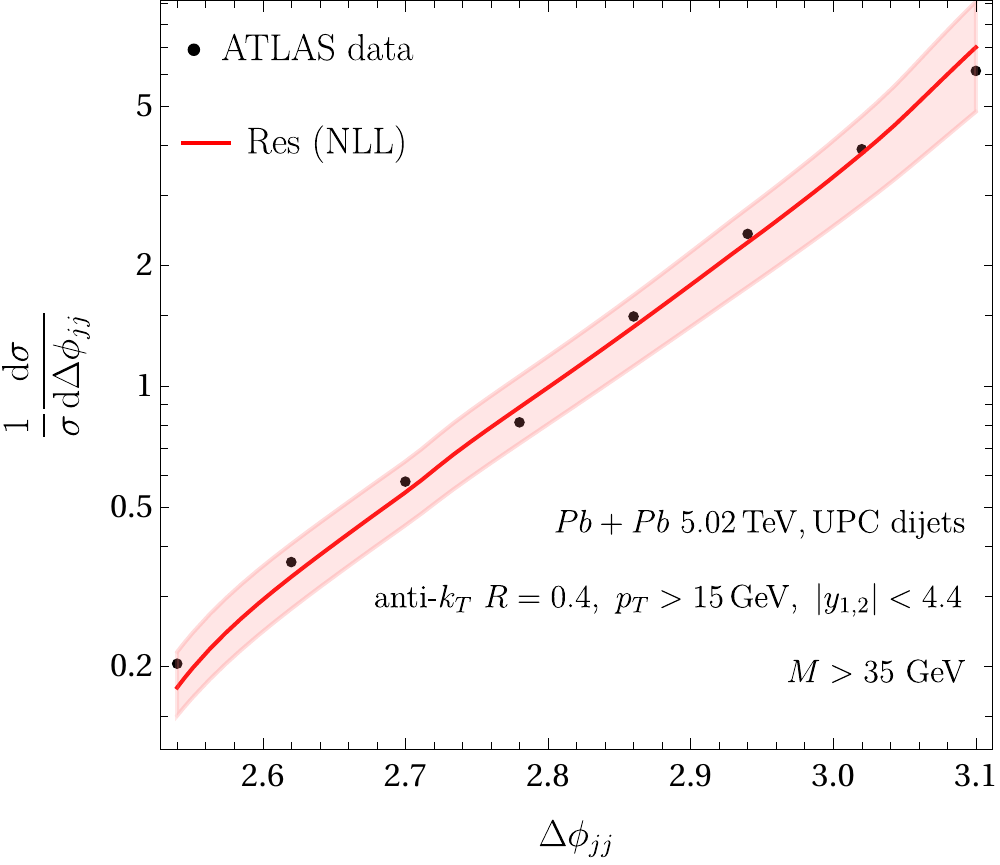}
  \caption{Comparison between theoretical calculations of the azimuthal decorrelation with the preliminary ATLAS data \cite{ATLAS:2022cbd}, where $\Delta\phi_{jj}$ is the azimuthal angle between two jets with highest transverse momenta. The theoretical and experimental distributions are normalized such that the area under the curve is equal to $1$.  The black dots are the ATLAS results, and the uncertainties of the data are smaller than the symbol size used in the plot. The solid red curve is the result with the scale choice in \eqref{eq:scales}, and the theoretical uncertainty for the red band is obtained from the variation of scales in the resummation formula. }\label{fig:theory-data}
\end{figure}

In the end we will compare our theoretical calculation of the azimuthal angular decorrelation $\Delta\phi_{jj}$ with the experimental results in \cite{ATLAS:2022cbd}, where we only consider the kinematic region where the jet pair is produced nearly back to back. In this region the contribution from multijets production is power suppressed, and we leave fixed-order pQCD calculations of multijets production for future studies.\footnote{Very recently, a new phenomenological code for automated fixed-order calculations of the photon-induced processes in UPCs is presented in \cite{Shao:2022cly}.}  In order to make a comparison, we calculate \eqref{eq:NLLres} with the same kinematic cuts as in \cite{ATLAS:2022cbd}
\begin{align}
  p_T>15\,{\rm GeV}, ~~~ |y_{1,2}|<4.4, ~~~ M>35\,{\rm GeV}, 
\end{align} 
and the jet radius $R=0.4$. As is shown in figure \ref{fig:theory-data} the red curve is the theoretical predictions with the scale choice in \eqref{eq:scales}, and the error bands are shown as the shaded regions where we consider the theoretical uncertainty from the scales. Specifically, we perform a variation of $\mu_h$, $\mu_j$, and $\mu_b$ scales by a factor of $2$ about the central values, and then combine these uncertainties by taking the envelope. The black dots are the ATLAS results, where we find experimental errors are small and can hardly be seen in the plot. The theoretical and experimental distributions are normalized such that the area under the curve is equal to $1$. From the plot it is clear to see that our result is consistent with the experimental data. 

\section{Conclusion}\label{sec:conclusion}

In this paper we study the azimuthal decorrelation of dijet productions
from photon-photon fusions in UPCs, where the dijet is produced nearly
back-to-back in the transverse plane with respect to the beam direction.
We calculate the key measurement, coplanarity distribution, which
is the distribution of the azimuthal angle $\Delta\phi_{jj}$
between the leading and the subleading jet. The azimuthal decorrelation
is produced via both the non-trivial transverse momentum distribution
of the incoming photon flux and QCD radiations from the outgoing
partons, which leads to the factorized formula \eqref{eq:fac_mom},
namely the convolution of the initial state and the final state. The
initial quark-antiquark pairs production is performed by the hard scattering of
the diphoton associated with the EPA photon spectrum. In addition,
the final-state radiations are described by the factorization and resummation
formula at NLL accuracy within the SCET framework. 

We vary the scales in resummation formula and obtain a theoretical
band of the normalized coplanarity distribution. Figure \ref{fig:theory-data}
shows a good agreement with the ATLAS experimental data in the nearly
back-to-back region, where we note that the \textsc{Pythia}8 Monte Carlo event generator
simulates a noticeably narrower $\Delta\phi_{jj}$ distribution
compared to the data~\cite{ATLAS:2022cbd}. However, both our theoretical calculations and
the \textsc{Pythia}8 simulation severely underpredict the total cross section
of the experiment (roughly by a factor of $10$), which implies some
missing progresses (e.g.~diffractive dijet photoproductions) in these
UPC $0n0n$ events to be studied in the future. We argue that these additional
dijet photoproductions may enhance the dijet production yields, but
should barely change the azimuthal angular distribution shape since it
is dominated by the final-state radiations.

\acknowledgments
C.Z.~is supported by the National Science Foundations of China under Grant No. 12147125. D.Y.S.~is supported by the National Science Foundations of China under Grant No.  12275052 and the Shanghai Natural Science Foundation under Grant No. 21ZR1406100.

\appendix

\section{Soft, collinear-soft functions and the collinear anomaly}\label{app:one-loop-funs}

The operator definition of the soft function can be found in \cite{Chien:2022wiq}, and here we summarize its one-loop results. The NLO soft function reads
\begin{align}\label{eq:softnlo}
  \tilde S(&b_x,y_1,y_2,\mu,\nu) =  1+ 2\,C_F g_s^2\tilde{\mu}^{2\epsilon} \int \frac{\mathrm{d}^dk}{(2\pi)^{d-1}}\delta(k^2)\theta(k^0) \left(\frac{\nu}{2k^0}\right)^\eta \frac{n_1 \cdot n_2}{n_1 \cdot k \, k \cdot n_2}e^{-i b_x k_x} \notag \\
  & = 1 + \frac{\alpha_s}{4\pi} C_F \left[ -\left(\frac{2}{\eta} + \ln \frac{\nu^2 n_1\cdot n_2}{2\mu^2}\right)\left(\frac{4}{\epsilon}+ 4 \ln \frac{\mu^2b_x^2}{b_0^2}\right) + \frac{4}{\epsilon^2} -2 \ln^2 \frac{\mu^2b_x^2}{b_0^2} - \frac{\pi^2}{3} \right],
\end{align}
with $\tilde{\mu}^{2} = \mu^2e^{\gamma_E}/(4\pi)$ and $b_0 = 2e^{-\gamma_E}$. Here the rapidity regulator is given in \eqref{eq:rap-reg}. In order to evaluate the one-loop collinear-soft function $U_i$, it is convenient to expand the integrated momentum $k^\mu$ along the light-like reference vector $n_i^\mu$. Explicitly, we have
\begin{align}
  k^\mu = \frac{n_i^\mu}{2}\bar n_i\cdot k + \frac{\bar n_i^\mu}{2} n_i\cdot k + k_\perp^\mu.
\end{align}
Then the one-loop collinear-soft function is written as
\begin{align}\label{eq:coftnlo}
  \tilde U_i&(b_\perp,y_i,\mu,\nu) =  1+ 2\,C_F g_s^2\tilde{\mu}^{2\epsilon} \int \frac{\mathrm{d}^dk}{(2\pi)^{d-1}}\delta(k^2)\theta(k^0) \left(\frac{\nu}{\bar n_i\cdot k}\right)^\eta \frac{n_i \cdot \bar n_i}{n_i \cdot k \, k \cdot \bar n_i} \notag \\
  & \times \left\{e^{-i \bm b_\perp \cdot \bm k_\perp} \theta\left[ \frac{n_i\cdot k}{\bar n_i\cdot k} - \left(\frac{R}{2\cosh y_i}\right)^2 \right] + \theta\left[ \left(\frac{R}{2\cosh y_i}\right)^2 -\frac{n_i\cdot k}{\bar n_i\cdot k}  \right] \right\} \notag \\
  & = 1 + \frac{\alpha_s}{4\pi} C_F \left[ \left(\frac{2}{\eta} + \ln \frac{\nu^2 R^2}{4\mu^2\cosh^2y_i}\right)\left(\frac{2}{\epsilon}+ 2 \ln \frac{\mu^2b_\perp^2}{b_0^2}\right) - \frac{2}{\epsilon^2} + \ln^2 \frac{\mu^2b_\perp^2}{b_0^2} + \frac{\pi^2}{6} \right],
\end{align}
where the rapidity regulator is also expanded at the leading power as suggested in \eqref{eq:coft-mode}. The collinear-soft mode describes low energy radiations emitted from the collinear partons in the jet at an angle $\theta\sim R$, and the step functions in the second line indicate that only emissions outside the jet contribute to transverse momentum imbalance at one loop. 

As shown in \eqref{eq:collinear_anomaly}, after combing soft and collinear-soft functions the jet radius dependence would be factorized out. At order $\alpha_s$, using \eqref{eq:softnlo} and \eqref{eq:coftnlo}, we find 
\begin{align}\label{eq:UUS-NLO}
   \tilde S (b_x, y_1, y_2,\mu,\nu) & \tilde U_{1}(b_x,y_1,\mu,\nu) \tilde U_{2}(b_x,y_2,\mu,\nu) \notag \\
  &  =1+C_{F}\frac{\alpha_{s}}{\pi} \bigg[\ln R^2 - \ln (2+2\cosh\Delta y)  \bigg]\left(\frac{1}{\epsilon}+\ln\frac{b_x^{2}\mu^{2}}{b_{0}^{2}}\right) + \mathcal{O}(\alpha_s^2),
\end{align}
where the poles in $\eta$ and scale $\nu$ dependence in the NLO terms cancel out. Since we consider $R\ll 1$ in \eqref{eq:fac}, the logarithm of $R$ indicates that a complete separation of scales was not achieved. In order to carry out the resummation of the logarithmic of the jet radius $R$, one obtain the refactorized formula in \eqref{eq:collinear_anomaly}. At one-loop order we find $\overline{\rm MS}$ renormalized result of \eqref{eq:UUS-NLO} agrees with \eqref{eq:collinear_anomaly} using the exponent $F_{q\bar q}$ in \eqref{eq:Fqq} and the reminder function $W$ in \eqref{eq:W-fun}. 

In the above calculation, we evaluate both the one-loop soft and collinear-soft functions in the lab frame where rapidities $y_{1,2}$ of dijet are arbitrary, and obtain general expansions of them. However, since the product of the soft and the collinear-soft functions are boost invariant along the beam axis as given in \eqref{eq:fac}, they can also be simultaneously calculated in the center-of-mass frame of two incoming photons, which would give the same result as in \eqref{eq:UUS-NLO} after combining them together.

\section{Anomalous dimension}\label{app:anoma_dim}
The QCD $\beta$-function and the cusp and non-cusp anomalous dimensions
in the $\overline{\textrm{MS}}$ renormalization scheme are expanded
as
\begin{equation}
\beta(\alpha_{s})=-2\alpha_{s}\sum_{n=0}^{\infty}\beta_{n}\left(\frac{\alpha_{s}}{4\pi}\right)^{n+1},\quad\gamma(\alpha_s)=\sum_{n=0}^{\infty}\gamma_{n}\left(\frac{\alpha_{s}}{4\pi}\right)^{n+1}.
\end{equation}
The two-loop coefficients of the $\beta$-function and the cusp anomalous
dimensions, and the one-loop coefficient of the non-cusp anomalous
dimensions read,
\begin{equation}
\beta_{0}=\frac{11}{3}C_{A}-\frac{4}{3}T_{F}n_{f},\quad\beta_{1}=\frac{34}{3}C_{A}^{2}-\frac{20}{3}T_{F}C_{A}n_{f}-4T_{F}C_{F}n_{f},
\end{equation}
\begin{equation}
\gamma_{0}^{\text{cusp }}=4,\quad\gamma_{1}^{\text{cusp }}=\left(\frac{268}{9}-\frac{4\pi^{2}}{3}\right)C_{A}-\frac{80}{9}T_{F}n_{f},\quad\gamma_{0}^{q}=-3C_{F},
\end{equation}
with $T_{F}=1/2,\:C_{A}=3,\:C_{F}=4/3,\:n_{f}=5.$
\bibliographystyle{JHEP}
\bibliography{jet.bib}

\end{document}